\documentclass[12pt]{article}
\usepackage{sw20lart}

\input{tcilatex}
\begin{document}

\author{Emilio Santos \\
Departamento de F\'{i}sica. Universidad de Cantabria. Santander. Spain}
\title{Mathematical and physical meaning of the Bell inequalities.}
\date{March, 27, 2016 }
\maketitle

\begin{abstract}
It is shown that the Bell inequalities are closely related to the triangle
inequalities involving distance functions amongst pairs of random variables
with values $\left\{ 0,1\right\} $. A hidden variables model may be defined
as a mapping between a set of quantum projection operators and a set of
random variables. The model is noncontextual if there is a joint probability
distribution. The Bell inequalities are necessary conditions for its
existence. The inequalities are most relevant when measurements are
performed at space-like separation, thus showing a conflict between quantum
mechanics and local realism (Bell's theorem). The relations of the Bell
inequalities with contextuality, Kochen-Specker theorem, and quantum
entanglement are briefly discussed.

\end{abstract}

\section{ Introduction}

More than fifty years have elapsed since John Bell derived his celebrated
inequalities\cite{Bell64}, \cite{Bell87}. These inequalities are fulfilled
by all classical theories but are violated by quantum predictions for some
quantum states. Thus they prove the impossibility of supplementing quantum
mechanics with local hidden variables, a result known as Bell's theorem.
This is popularly stated saying that local realism is incompatible with
quantum mechanics. Bell's work had a relevant impact on the foundations of
quantum mechanics. In particular, it has produced a renewed interest in the
study of quantum entanglement. It has also led to increasingly refined
experiments and it has been a stimulus for the development of quantum
information theory. In the last year Bell's theorem has received new
attention due to the fact that loophole-free tests of the inequalities have
been performed for the first time\cite{Hensen}, \cite{Shalm}, \cite{Giustina}%
.

In a wide sense the name ``Bell inequality'' is applied to any inequality
that is a necessary condition for the existence of local realisitc (or local
hidden variables models). For a recent review see\cite{Brunner}. In this
paper I will be concerned with a different, although related, definition of
Bell inequalities, namely quadrilateral inequalities involving distances
defined in (classical) probability theory for sets of dichotomic random
variables. The fulfillement, or not, of the inequalities for similar
distances defined for expectation values of projection operators in a
Hilbert space allows us to know whether some subsets of operators may be
represented, or not, in terms of (classical) probability theory. This
connection is stressed in our derivation of the inequalities in section 2.

Actually, those properties are related to the distributivity of classical
logic. Indeed, as Birkhoff and von Neumann pointed out\cite{Hooker}, the
essential difference beween classical and quantum physics appears already at
the level of the logic (or propositional calculus), which is distributive in
the classical case but not in the quantum case. This relation of the Bell
inequalities with the distributivity of the logic has been discussed since
long ago\cite{Santos86}, \cite{Pitowski}, \cite{Beltrametti}, \cite{Dvur}, 
\cite{Pulmannova3}, \cite{Pykacz}.

The physical implications of the Bell inequalities will be briefly discussed
in section 3. However, I shall touch but slightly on the experimental tests
of the inequalities and not at all on the applications to quantum
information theory. The main purpose of the section will be to comment on
the relevance of the inequalities for the interpretation of quantum
mechanics, in particular the question whether a realistic interpretation is
possible. There are a lot of papers and many books concerned with this
question, that will not be commented here (see, e.g. \cite{Auletta}, \cite
{Laloe}, \cite{Redhead}, \cite{d'Espagnat1}, \cite{d'Espagnat2}). The
subject is still controversial\cite{Englert}.

\section{Mathematical content}

From the mathematical point of view the Bell inequalities are necessary
conditions for the existence of a joint probability distribution associated
to a set of projection operators (projectors) and a density operator on a
Hilbert space. A relevant result is that, if all projectors of the set
commute with each other, then a joint probability distribution exists and
all Bell inequalities hold true for any density operator. Although this
section deals with mathematical properties, and therefore it is independent
of any physical theory, I will use the widely known language of quantum
mechanics. In particular \textit{in this section} I shall name
``observable'' any self-adjoint operator in the Hilbert space.

\subsection{Probability distribution of an observable}

For any pair $\left\{ \hat{A},\hat{\rho}\right\} $ of an observable and a
density operator it is possible to define the probability density, $f\left(
a\right) ,$ of a random variable, $a$, associated with the observable. In
fact the standard rule about the expectation value, $\left\langle \hat{A}%
^{n}\right\rangle ,$ of the n'th power of an observable $\hat{A}$ in the
state given by the density operator $\hat{\rho}$ allows getting the
characteristic function, $C(\zeta ),$ of the associated probability
distribution, that is 
\[
\left\langle \hat{A}^{n}\right\rangle =Tr\left[ \hat{A}^{n}\hat{\rho}\right]
\Rightarrow C(\zeta )=Tr\left[ \exp \left( i\zeta \hat{A}\right) \hat{\rho}%
\right] , 
\]
where $Tr(\hat{x})$ means the trace of the operator $\hat{x}$. The
probability density is obtained by Fourier transform, that is 
\begin{eqnarray}
f\left( a\right) &=&\frac{1}{2\pi }\int d\zeta \exp \left( -i\zeta a\right)
C(\zeta )  \nonumber \\
&=&\frac{1}{2\pi }\int d\zeta \exp \left( -i\zeta a\right) Tr\left[ \exp
\left( i\zeta \hat{A}\right) \hat{\rho}\right] .  \label{1}
\end{eqnarray}

\emph{Example 1: Observable }$\hat{A}$\emph{\ having discrete nondegenerate
spectrum and }$\hat{\rho}$\emph{\ being a pure state, }that is\emph{\ } 
\[
\hat{A}=\sum_{j}a_{j}\mid \psi _{j}\rangle \langle \psi _{j}\mid ,\hat{\rho}%
=\mid \phi \rangle \langle \phi \mid . 
\]
Assuming that the eigenfunctions $\left\{ \mid \psi _{j}\rangle \right\} $
form an orthonormal basis we have 
\begin{eqnarray*}
\exp \left( i\zeta \hat{A}\right) &=&\sum_{k}\frac{\left( i\zeta
\sum_{j}a_{j}\mid \psi _{j}\rangle \langle \psi _{j}\mid \right) ^{k}}{k!} \\
&=&\sum_{k}\frac{\left( i\zeta \right) ^{k}}{k!}\sum_{j}(a_{j})^{k}\mid \psi
_{j}\rangle \langle \psi _{j}\mid =\sum_{j}\exp \left( i\zeta a_{j}\right)
\mid \psi _{j}\rangle \langle \psi _{j}\mid ,
\end{eqnarray*}
whence, taking eq.$\left( \ref{1}\right) $ into account, the probability
distribution is 
\begin{eqnarray*}
f\left( a\right) &=&\frac{1}{2\pi }\sum_{j}\int d\zeta \exp \left( -i\zeta
a+i\zeta a_{j}\right) \langle \phi \mid \psi _{j}\rangle \langle \psi
_{j}\mid \phi \rangle \\
&=&\sum_{j}\delta \left( a-a_{j}\right) p_{j},\;p_{j}\equiv \left| \langle
\phi \mid \psi _{j}\rangle \right| ^{2}.
\end{eqnarray*}
This means that the distribution is discrete and it associates the
probability $\left| \langle \phi \mid \psi _{j}\rangle \right| ^{2}$ to the
eigenvalue $a_{j}$.

\emph{Example 2. \^{A} is a projector and }$\hat{\rho}$ \emph{arbitrary}
(not necessarily representing a pure state). $\hat{A}$ fulfils 
\[
\hat{A}=\hat{A}^{2}\Rightarrow \hat{A}^{n}=\hat{A}, 
\]
for any natural number $n$. Then we have 
\begin{eqnarray*}
f\left( a\right) &=&\frac{1}{2\pi }\sum_{j}\int d\zeta \exp \left( -i\zeta
a\right) Tr\left[ \hat{\rho}\exp \left( i\zeta \hat{A}\right) \right] \\
&=&\frac{1}{2\pi }\sum_{j}\int d\zeta \exp \left( -i\zeta a+i\zeta \right)
Tr\left( \hat{\rho}\hat{A}\right) =\delta \left( a-1\right) Tr\left( \hat{%
\rho}\hat{A}\right) ,
\end{eqnarray*}
which means that $Tr\left( \hat{\rho}\hat{A}\right) =p(a)$ is the
probability that the projector takes the value $1$.

\subsection{Joint distribution for several observables}

Generalizing eq.$\left( \ref{1}\right) $ we might define the joint
probability distribution of two observables, $f\left( a,b\right) ,$ as
follows 
\begin{equation}
f\left( a,b\right) =\frac{1}{4\pi ^{2}}\int d\zeta \int d\chi \exp \left(
-i\zeta a-i\chi b\right) Tr\left[ \exp \left( i\zeta \hat{A}+i\chi \hat{B}%
\right) \hat{\rho}\right] .  \label{1.1}
\end{equation}
If the observables $\hat{A}$ and $\hat{B}$ commute with each other, the
function $f\left( a,b\right) $ is indeed the desired joint distribution and
it has the properties of a classical joint distribution, as shown in the
following. In fact in this case there is a complete set of orthonormal
simultaneous eigenvectors of the two observables, that we will label $%
\left\{ \mid \phi _{j}\rangle \right\} $ (assuming for simpliticity that the
set is discrete). Thus we may use the resolution of the identity operator $%
\hat{I}$ in terms of these eigenvectors in order to define the trace, i. e. 
\[
\hat{I}=\sum_{j}\mid \phi _{j}\rangle \langle \phi _{j}\mid ,Tr\hat{M}%
=\langle \phi _{j}\mid \hat{M}\mid \phi _{j}\rangle . 
\]
Hence eq.$\left( \ref{1.1}\right) $ becomes, after some algebra, 
\begin{equation}
f\left( a,b\right) =\sum_{j}P_{j}\rho _{aj}(a)\rho _{bj}(b),P_{j}=\langle
\phi _{j}\mid \hat{\rho}\mid \phi _{j}\rangle ,  \label{1.2}
\end{equation}
where $\rho _{aj}(a)$ is the probability distribution of $a$ in the state $j$%
, as given by eq.$\left( \ref{1}\right) .$ Eq.$\left( \ref{1.2}\right) $ has
the same form as a classical joint probability distribution written in terms
of the probability distributions of the individual variables $a$ and $b$. In
particular if the sum in $j$ contains only one term, the distributions $\rho
_{aj}(a)$ and $\rho _{bj}(b)$ are uncorrelated.

In sharp contrast if $\hat{A}$ and $\hat{B}$ do not commute the function $%
f\left( a,b\right) $ defined by eq.$\left( \ref{1.1}\right) $ may not be
semidefinite positive. For instance in a system consisting of a single
particle (without spin) in one dimension, the position and momentum
observables, $\hat{x}$ and $\hat{p},$ do not commute and eq.$\left( \ref{1.1}%
\right) $ leads to 
\[
W\left( x,p\right) =\frac{1}{4\pi ^{2}}\int d\zeta \int d\chi \exp \left(
-i\zeta x-i\chi p\right) Tr\left[ \exp \left( i\zeta \hat{x}+i\chi \hat{p}%
\right) \hat{\rho}\right] , 
\]
which is the Wigner function associated to the state $\hat{\rho}.$ As is
well known the Wigner function is not always positive. We might try other
choices for the ordering of the operators $\hat{x}$ and $\hat{p},$ but none
fully satisfactory has been found. Thus the question arises, is it always
possible to find a joint (positive semidefinite) probability distribution
for several observables and a given density operator?. The answer is
affirmative if all observables commute, but negative in general. In order to
prove that assertion let us start studying joint probability distributions
in the mathematical theory of probability.

\subsection{Joint distribution for several dichotomic random variables}

What I will present in the following is well kown for any mathematical \emph{%
measure, }in particular a probability distribution, defined on a set. In
mathematics a measure on a set is a mapping of subsets on nonnegative
numbers, with some topological restrictions on the subsets if the set is not
finite. But for our purposes it is enough to study probability distributions
on a finite set $\left\{ a_{j},j=1,2,...n\right\} $ of random variables
having values $\left\{ 0,1\right\} $.

The question that I will try to answer is the following. Given the
probabilities $\left\{ p_{1}\left( a_{j}\right) \right\} $ that $a_{j}=1$,
and the probabilities $\left\{ p_{2}\left( a_{j}a_{k}\right) \right\} $ that 
$a_{j}=a_{k}=1$ for all $j,k$, we want to know whether these probabilities
are the marginals of some joint probability distribution for all the
variables. A well known result of probability theory is the following lemma: 
\emph{A necessary condition for the existence of a joint probability
distribution is that all triangle inequalities of the form } 
\begin{equation}
d(a_{j},a_{k})+d(a_{k},a_{l})\geq d(a_{j},a_{l})  \label{10}
\end{equation}
\emph{hold true}. The ``distance function'' $d(a_{j},a_{k})$ is defined by 
\begin{equation}
d(a_{j},a_{k})=p(a_{j}\bar{a}_{k})+p(\bar{a}_{j}a_{k})=p_{1}\left(
a_{j}\right) +p_{1}\left( a_{k}\right) -2p_{2}\left( a_{j}a_{k}\right) ,
\label{11}
\end{equation}
where $\bar{a}_{j}=1-a_{j}$ is the variable that takes the value $1(0)$ when 
$a_{j}$ takes the value $0(1)$. The distance function fulfils 
\begin{equation}
0\leq d(a_{j},a_{k})\leq 1,d(a_{j},a_{j})=0,d(a_{j},\bar{a}_{j})=1,
\label{12}
\end{equation}
We see that the distance is zero when the two variables are maximally
correlated (they might be considered the same variable) and it is unity if
they are maximally anticorrelated. The proof of the lemma is easy. We
consider 3 random varibles $\left\{ a,b,c\right\} $ with values $\left\{
0,1\right\} $ and use a Venn diagram where the random variables are
represented by circles and the areas are proportional to the probabilities.
Thus it is trivial to check that 
\[
p(a\bar{b})+p(\bar{a}b)+p(a\bar{c})+p(\bar{a}c)\geq p(b\bar{c})+p(\bar{b}c), 
\]
which leads to a triangle inequality similar to eq.$\left( \ref{10}\right) $.

In the following I show that the triangle inequalities eq.$\left( \ref{10}%
\right) $ are closely related to the Bell inequalities\cite{Bell64}.

\subsection{Random variables representations}

Let us consider a finite set, $P\equiv \left\{ \hat{a}_{j}\right\} ,$ of
projectors (with eigenvalues $1$ and $0$) and a density operator $\hat{\rho}$%
. A set $R\equiv \left\{ a_{j}\right\} $ of random variables with values $%
\left\{ 0,1\right\} $ is here defined as a ``random variables
representation'' (RVR) of the pair $\left\{ P,\hat{\rho}\right\} $ if there
is an one to one mappping $P\rightarrow R$ such that, for any subset of $P$
involving only projectors that commute with each other, the following
equality holds true 
\begin{equation}
Tr\left[ \hat{\rho}\left( \hat{a}_{j}\hat{a}_{k}...\hat{a}_{l}\right)
\right] =p\left( a_{j}a_{k}...a_{l}\right) ,  \label{16}
\end{equation}
where $p\left( a_{j},a_{k}...a_{l}\right) $ is the probability that all
variables $\left\{ a_{j},a_{k}...a_{l}\right\} $ take the value $1$. We
assume that both sets $P$ and $R$ are complete in the sense that if $\hat{a}%
_{j}\in P$ also $\hat{I}-\hat{a}_{j}\in P,$ $\hat{I}$ being the identity
operator. Hence if $a_{j}\in R$ also $\bar{a}_{j}\in R.$

The quantities on the right side of eq.$\left( \ref{16}\right) $ have all
the properties of probabilities. In fact they are positive or zero and the
marginals are consistent with the mapping $P\rightarrow R$, that is 
\begin{eqnarray*}
p\left( a_{k}...a_{l}\right) &=&p\left( a_{j}a_{k}...a_{l}\right) +p\left( 
\bar{a}_{j}a_{k}...a_{l}\right) \\
&=&Tr\left[ \hat{\rho}\left( \hat{a}_{j}\hat{a}_{k}...\hat{a}_{l}\right)
\right] +Tr\left[ \hat{\rho}\left( (\hat{I}-\hat{a}_{j})\hat{a}_{k}...\hat{a}%
_{l}\right) \right] \\
&=&Tr\left[ \hat{\rho}\left( \hat{a}_{k}...\hat{a}_{l}\right) \right] .
\end{eqnarray*}
In conclusion \emph{a\ ``random variables representation'' exists for any
pair }$\left\{ P,\hat{\rho}\right\} $.

\emph{Example: }Let us consider four projectors $\left\{ \hat{a}_{1},\hat{b}%
_{1},\hat{a}_{2},\hat{b}_{2}\right\} $ such that $\hat{a}_{j}$ commutes with 
$\hat{b}_{k}$ for any $j,k\in \left\{ 1,2\right\} $, but neither $\hat{a}%
_{1} $ commutes with $\hat{a}_{2}$ nor $\hat{b}_{1}$ with $\hat{b}_{2}.$
Thus eqs.$\left( \ref{16}\right) $ allow defining 24 probabilities, namely 
\begin{equation}
p(a_{j}),p(b_{k}),p(\bar{a}_{j}),p(\bar{b}_{k}),p(a_{j}b_{k}),p(\bar{a}%
_{j}b_{k}),p(a_{j}\bar{b}_{k}),p(\bar{a}_{j}\bar{b}_{k}).  \label{17}
\end{equation}
But there are many probabilities which are not defined by eqs.$\left( \ref
{16}\right) ,$ namely those involving two or more noncommuting operators.
For instance the following probabilities cannot be derived that way 
\begin{equation}
p(a_{1}a_{2}),p(b_{1}b_{2}),p(a_{1}a_{2}b_{k}),p(b_{1}b_{2}a_{j}),p(a_{1}a_{2}b_{1}b_{2}).
\label{17a}
\end{equation}

There are two different kinds of random variables representation. \emph{A
RVR\ of the pair }$\left\{ P,\hat{\rho}\right\} $\emph{\ is ``complete'' if
there is a joint probability distribution for all random variables of R such
that the probabilities given by eqs.}$\left( \ref{16}\right) $ \emph{are
marginals of the joint distribution. It is incomplete if such a joint
probability does not exist.}

In the above example a complete RVR would require the existence of the 16
probabilities 
\begin{equation}
p(a_{1}a_{2}b_{1}b_{2}),p(a_{1}a_{2}b_{1}\bar{b}_{2}),p(a_{1}a_{2}\bar{b}%
_{1}b_{2}),...p(\bar{a}_{1}\bar{a}_{2}\bar{b}_{1}\bar{b}_{2}),  \label{17b}
\end{equation}
whence all other probabilities could be obtained as marginals. However some
of these could not be derived from eqs.$\left( \ref{16}\right) ,$ for
instance this would be the case for the probabilities eqs.$\left( \ref{17a}%
\right) $ in our example. On the other hand it may be that probabilities eqs.%
$\left( \ref{17b}\right) $ do not exist with the condition that all
probabilities derivable from eqs.$\left( \ref{16}\right) $ are marginals of
these eqs.$\left( \ref{17b}\right) $. In this case the RVR will not be
complete. The complete and incomplete RVR correspond to the socalled
noncontextual and contextual hidden variables theories in quantum mechanics,
respectivley, as will be clarified in the next section.

It is not difficult to generalize RVR to observables not necessarily
projectors. Actually any set of observables may be written (or accuratelly
approximated) in terms of projectors.

\subsection{Quadrilateral inequalities as tests of completeness}

An interesting task is to find conditions that allow to discover whether the
RVR of a pair $\left\{ P,\hat{\rho}\right\} $ is complete. In view of the
subsection 1.3 a necessary condition would be that all triangle inequalities
involving probabilities defined by eq.$\left( \ref{16}\right) $ are
fulfilled. However this criterion is useless. In fact for any 3 commuting
projectors the triangle inequality holds true either if the RVR is complete
or not. On the other hand if two projectors do not commute their distance
function cannot be defined because the joint probability is not given by eq.$%
\left( \ref{16}\right) .$ In this cases one of the sides of the triangle
inequality is unknown.

The solution to the problem is to consider quadrilateral inequalities as
follows. If the RVR is complete, then a joint probability is defined for any
3 random variables of the model. As a consequence, choosing any four
variables $\left\{ a_{1},a_{2},b_{1},b_{2}\right\} $ the following two
triangle inequalities hold true 
\[
d(a_{1},b_{1})\leq d(a_{1},a_{2})+d(a_{2},b_{1}),d(a_{1},a_{2})\leq
d(a_{1},b_{2})+d(b_{2},a_{2}), 
\]
although it may be that some of the distances cannot be obtained via eqs.$%
\left( \ref{16}\right) $ (but they might be obtained as marginals of the
assumed joint probability distribution as the RVR is complete). The addition
of these two inequalities gives the quadrilateral inequality 
\begin{equation}
d(a_{1},b_{1})\leq d(a_{1},b_{2})+d(b_{2},a_{2})+d(a_{2},b_{1}),  \label{15}
\end{equation}
which is usefull because all four distances may be obtained via eqs.$\left( 
\ref{16}\right) $. As a consequence a necessary condition for a RVR to be
complete is that all quadrilateral inequalities like eq.$\left( \ref{15}%
\right) $\emph{\ }hold true. The inequalities are not trivial if some of the
projectors associated with the variables do not commute. For instance, in
our example above the projectors $\hat{a}_{1}$ and $\hat{a}_{2}$ do not
commute.

\emph{The quadrilateral inequalities are closely related to the most typical
Bell inequalities}\cite{Bell64}, \cite{Bell87}. In fact, taking eq.$\left( 
\ref{11}\right) $ into account, $\left( \ref{15}\right) \ $may be written in
the form 
\begin{equation}
p\left( a_{1}\right) +p\left( b_{1}\right) \geq p(a_{1},b_{1})+p\left(
a_{2},b_{1}\right) +p\left( a_{1},b_{2}\right) -p\left( a_{2},b_{2}\right) ,
\label{CH}
\end{equation}
which is an inequality derived by Clauser and Horne\cite{CH} in 1974 from
different (but equivalent) assumptions as those used here. Most popular is
the inequality derived in 1969 by Clauser, Horne, Shimony and Holt (CHSH)%
\cite{CHSH} which is acually equivalent to eq.$\left( \ref{CH}\right) .$ In
fact, let us consider dichotomic variables $\left\{ A_{j},B_{k}\right\} $
with values $\left\{ -1,1\right\} ,$ related to the variables $\left\{
a_{j},b_{k}\right\} $, with values $\left\{ 0,1\right\} ,$ by means of 
\begin{equation}
a_{j}=\frac{1}{2}\left( A_{j}+1\right) ,b_{k}=\frac{1}{2}\left(
B_{k}+1\right) .  \label{ab}
\end{equation}
Now for a random variable $a$ with values $\left\{ 0,1\right\} $ the
probability $p(a)$ of taking the value $1$ is identical to the expectation
value $\left\langle a\right\rangle $. Thus the inequality eq.$\left( \ref{CH}%
\right) $ may be rewritten 
\[
\left\langle a_{1}\right\rangle +\left\langle b_{1}\right\rangle \geq
\left\langle a_{1}b_{1}\right\rangle +\left\langle a_{2}b_{1}\right\rangle
+\left\langle a_{1}b_{2}\right\rangle -\left\langle a_{2}b_{2}\right\rangle
, 
\]
which, taking eq.$\left( \ref{ab}\right) $ into account, leads to 
\begin{eqnarray*}
\frac{1}{2}\left\langle A_{1}+1\right\rangle +\frac{1}{2}\left\langle
B_{1}+1\right\rangle &\geq &\frac{1}{4}\left\langle \left( A_{1}+1\right)
\left( B_{1}+1\right) \right\rangle +\frac{1}{4}\left\langle \left(
A_{2}+1\right) \left( B_{1}+1\right) \right\rangle \\
&&+\frac{1}{4}\left\langle \left( A_{1}+1\right) \left( B_{2}+1\right)
\right\rangle -\frac{1}{4}\left\langle \left( A_{2}+1\right) \left(
B_{2}+1\right) \right\rangle .
\end{eqnarray*}
Hence simple algebra gives 
\begin{equation}
2\geq \left\langle A_{1}B_{1}\right\rangle +\left\langle
A_{2}B_{1}\right\rangle +\left\langle A_{1}B_{2}\right\rangle -\left\langle
A_{2}B_{2}\right\rangle ,  \label{CHSH}
\end{equation}
which is the CHSH inequality\cite{CHSH}.

I shall point out that in papers dealing with the CHSH inequality it is
frequent to call ``correlation'' the expectation value of a product of
observables like $\left\langle AB\right\rangle .$ The name does not agree
with the standard one in mathematical statistics, where the correlation
between two random variables, $A$ and $B,$ is usually defined by the
dimensionless quantity 
\begin{equation}
Corr(A,B)\equiv \frac{\left\langle AB\right\rangle -\left\langle
A\right\rangle \left\langle B\right\rangle }{\sqrt{\left\langle
A^{2}\right\rangle -\left\langle A\right\rangle ^{2}}\sqrt{\left\langle
B^{2}\right\rangle -\left\langle B\right\rangle ^{2}}}.  \label{corr}
\end{equation}
An inequality similar to CHSH with the correlation eq.$\left( \ref{corr}%
\right) $ substituted for the expectation of the product, $\left\langle
AB\right\rangle ,$ may be violated by classical (hidden variables) models.
This fact has lead some authors to misunderstand, and criticize, Bell\'{}s
work.

\section{Physical implications}

The physical meaning of the Bell inequalities has been widely discussed and
a detailed survey is beyond the scope of this paper. Here I will give a
summary of the fundamental ideas. The interested reader may consult the
review by Brunner et al.\cite{Brunner}. At a difference with the previous
(mathematical) section, here it is necessary to distinguish between an
``observable'' of a physical system, say $A$, and the ``operator''
associated to it in the Hilbert space formalism of quantum mechanics, say $%
\hat{A}$.

\subsection{Observables and elements of reality}

\subsubsection{From EPR to hidden variables}

The starting point of Bell\'{}s work was the celebrated article by Einstein,
Podolsky and Rosen (EPR)\cite{EPR}. Indeed the title of the pioneer Bell
publication made reference to that paper\cite{Bell64}. EPR article begins
with what may be seen as a definition of \textit{epistemological realism, }%
that is\textit{\ }the conditions for\textit{\ a realistic interpretation of
physics}: ``Any serious consideration of a physical theory must take into
account the distinction between the objective reality, which is independent
of any theory, and the physical concepts with which the theory operates.
These concepts are intended to correspond with the objective reality, and by
means of these concepts we picture this reality to ourselves''\cite{EPR}.
Then the authors present a state of a pair of particles entangled in
positions and momenta. They stress the quantum prediction that, even if the
particles are well separated, a measurement of position (momentum) on the
first particle allows an instantaneous knowledge of the position (momentum)
of the second one. Thus we should attach to the second particle different
quantum states without in any way making a physical action on it. The
authors conclude that either there is an action at a distance (nonlocality)
or quantum mechanics associates two different quantum wavefunctions to the
same actual state of the second particle. Thus the EPR paper proves that
quantum mechanics is either nonlocal or incomplete, and the authors
considered the second alternative more plausible. Bell showed that if
quantum mechanics is incomplete (and it admits hidden variables) it is also
incompatible with local realism (Bell's theorem). Therefore the question of
completeness is irrelevant for the question of compatibility. In a soon
reply to EPR, Bohr supported completeness and a kind of wholeness of quantum
mechanics that implies nonlocality\cite{BohrEPR}. Thus Bell's theorem is
currently interpreted as a vindication of Bohr against Einstein.

Many physicists do not accept EPR realism, they believe that ``pictures of
the objective reality'' are irrelevant for physics where only the agreement
of the theoretical predictions with the empirical evidence is required.
However other physicists search for a clear picture of the reality, which is
not offered by the quantum formalism. A standard approach to a picture is
the introduction of the socalled ``hidden variables theories'', that might
reduce quantum theory to a stochastic theory with classical flavour. The
main achievement of Bell\'{}s work was to put constraints on the hidden
variables models that are possible, in particular proving the impossibility
of local models.

\subsubsection{Measurement in macro and microsystems}

Essential in any branch of natural science are the experiments, consisting
of just observations (as in astronomy) or preparations followed by
measurements (as in laboratory work). Thus it is worth to study the
differences between the experiments dealing with macroscopic and microscopic
systems.

The most popular difference is of course the fact the some quantities may
possess values only within a discrete set in microphysics. Hence the name 
\textit{quantum} introduced by Planck. However more important for the hidden
variables problem is the empirical fact that macroscopic systems are
perturbed but slightly by measurements, whilst in the microscopic domain the
state of the system is usally dramatically changed. One consequence is that
the uncertainties in the results of the measurements in the macroscopic
domain are usually small as compared with the result obtained. Thus \textit{%
it is assumed} that the perturbation may be reduced indefinitely and
therefore ignored in the formulation of classical theories, dealing with
macroscopic systems. The logical consequence is to attach the observable
quantities to the physical systems themselves, whence the belief that
measurements just revail intrinsic properties of the system under study.
Thus classical theories dispense with a theory of measurement. In fact the
process of measurement may be seen as just an interaction between the
measured system and the measuring setup, and therefore a process to be
studied within the theory itself.

In sharp contrast the measurement may give rise to a large perturbation in
quantum physics. A consequence of this fact is that, whilst the simultaneous
measurement of two observables on a macroscopic system perturb but slightly
each other, the perturbation may be very relevant in joint measurements of
microscopic systems. Therefore in classical theories \textit{we may assume}
that all observables of a system are \textit{compatible}, in the sense that
they may be measured simultaneously. (In macroscopic systems there are
examples of obsevables that cannot be measured simultaneously, but they are
scarce and irrelevant for the formulation of classical theories). In sharp
contrast \textit{for microsopic systems there are incompatible observables}
that cannot be measured with the same experimental setup. The quantum
mechanical formalism incorporates this fact in a fundamental form, namely
two observables are compatible if and only if the associated operators
commute.

\subsubsection{Incompatible observables}

The existence of incompatible measurements in microscopic systems leads
logically to the possibility that the results of joint measurements of two
(or several) observables depend on the context. For instance, the
measurement of the position of a particle requires a different experimental
context than the measurement of momentum. Therefore the joint measurement of
position and spin requires a context different from the one for a joint
measurement of momentum and spin. Thus we might expect that the spin
measurement (in the same state, that is after the same preparation
procedure) gives different results in the two cases. This is not so in our
example (in fact the result for the spin measurement would the same in both
cases) but different results are actually obtained in more complex examples,
involving measurements of more than two observables. The property is a
consequence of the quantum formalism and it is formally stated by the
Kochen-Specker theorem (see below).

\subsubsection{Are there objective properties?}

As said above in classical physics it is assumed that measurements just
reveal properties of the physical systems, existing with independence of
observations. That is, in any actual (pure) state of a classical system all
observable quantities have a well defined value. A mixed state (or
statistical ensemble of systems) is associated to a joint probability
distribution of the observables. This is not true in quantum physics, where
it is not always possible to associate a joint probability distribution to
all observables in a specific quantum state (i.e. to a given preparation
procedure of the system). This fact has led some people to claim that, in
contrast with classical physics, quantum systems have no properties whenever
they are not measured. That is the observable properties ``are created'' or
``emerge'' as a result of measurements. This claim seemed bizarre to many
people including Einstein, who criticized it with the rhetorical sentence:
Is the moon there when nobody looks?.

In order to clarify the subject it is necessary to distinguish between the
objective properties or ``elements of reality''\cite{EPR}, that exist
independently of any observation, and the ``observable quantities'' which
appear only as a consequence of the observation or measurement. This is true
in both classical and quantum physics. The point may be seen with an
examples. In playing dice, every die has spots on the six faces with the
numbers 1 to 6. \textit{These are elements of reality}, always present in
the faces of the die. But if we play dice the value of \textit{our observable%
} is only one number per die, namely the one printed in the upper face when
the die becomes at rest on the table. The value of the observable, but not
the objective property, is ``created'' by the action of playing dice, i. e.
the experiment. In summary we should distinguish between objective
properties and observables both in classical and quantum physics. However
the connection between both is rather obvious in classical physics but less
clear in quantum physics.

I conclude that the differences between classical and quantum theories do
not provide sufficient support for the claim that objective properties (or
elements of reality) do not exist in quantum systems or that realistic
interpretations of quantum physics are not possible, that is interpretations
in terms of ``pictures of the reality''\cite{EPR}. These pictures might be
achieved by means of \textit{ontic }or\textit{\ hidden variables }models of
quantum systems, to be studied in the next subsection.

\subsection{Ontic models in classical and quantum physics}

Following the advise of the initial paragraph of the EPR paper we may assume
that a physical system is at any time in some ``ontic state'', that is ``a
real physical state not necessarily completely described by any theory (e.
g. quantum), but objective and independent of the observer''\cite{Spekkens}.
The observable properties of the system derive from the ontic state at the
time of observation\cite{FOS}.

The standard classical assumption that measurements just reveal existing
properties may be formalized stating that the observed result, $a$, depends
on the ontic state, $\lambda ,$ of the system and the measuring set up, $A,$
appropriate for a given observable quantity. That is the observed result, $a$%
, will be a function 
\begin{equation}
a=a\left( \lambda ,A\right) .  \label{Bellnonc}
\end{equation}
At this moment I stress that this relation offers a ``picture of the
objective reality'' as demanded by EPR, in the sense that it provides a
causal connection between system plus measuring setup and the result
obtained. In contrast the assumption that the results of the measurement
``emerge'' from the act of measurement would not provide any clear picture.
This is true specially if it is assumed that the randomness of the result is
``essential'' in the sense that it is not caused by the incomplete control
of all the parameters in the measurement. Eq.$\left( \ref{Bellnonc}\right) $
explains the randomness of the result from the incomplete control of either
the preparation procedure (leading to a statistical distribution of ontic
states, $\lambda )$ or the measuring setup, or both.

The assumption that the results of all observations on a system derive from
functions like $a\left( \lambda ,A\right) $, eq.$\left( \ref{Bellnonc}%
\right) ,$ would allow getting the joint probability of any set of
observables, $\left\{ A,B,...C\right\} .$ Without loss of generality we may
consider that the set consists of observables with values $\left\{
0,1\right\} ,$ because any observable may be defined in terms of yes-no
questions. In this case any function like $a\left( \lambda ,A\right) $ takes
on the value $1$ for some ontic states $\lambda $ and $0$ for other states.
Thus the joint probability distribution reduces to the knowledge of all
expectations of products of observables, that is 
\begin{equation}
\left\langle AB...C\right\rangle =\int f\left( \lambda \right) a\left(
\lambda ,A\right) b\left( \lambda ,B\right) ...c\left( \lambda ,C\right)
d\lambda .  \label{Bellhv}
\end{equation}
Here we have assumed that $\lambda $ is a numerical parameter (or a set of
parameters) attached to the ontic state so that the integral in $\lambda $
makes sense. The function $f\left( \lambda \right) $ gives the probability
distribution of the ontic states in cases where we use a statistical
ensemble of states (a ``mixed state'') due to incomplete information. For a
pure state the function $f\left( \lambda \right) $ is zero except for some
value of $\lambda $ (the function $f\left( \lambda \right) $ will have the
form of a Dirac's delta) which reduces the right side of eq.$\left( \ref
{Bellhv}\right) $ to a simple product (no integration is required).

\emph{The conclusion is that fixing the functions eqs.}$\left( \ref{Bellnonc}%
\right) $ \emph{for all possible observables of the system would determine
their joint probability distribution. }The reciprocal is also true. In fact,
it is enough to identify the parameter $\lambda $ with the set of values of
all observables.

The above construction, eq.$\left( \ref{Bellhv}\right) ,$ may be called
``ontic model''\cite{Spekkens} and extended, with some modifications, to
quantum physics. For some people the appeal to ontic states is a
philosophical (methaphysical) assumption that should not enter physics and
prefer to treat $\lambda $ as a parameter in a model for an experiment and
name it ``hidden variable''. Thus the construction may be called \textit{%
hidden variables model (HVM), }which is the common name in the context of
quantum physics.

Let us analyze whether HVM are possible in quantum mechanics. We consider a
quantum system (pure or mixed), a state given by the density operator $\hat{%
\rho},$ and the set of projection operators $\left\{ \hat{A},\hat{B}%
,...\right\} .$ The projectors, fulfilling $\hat{A}^{2}=\hat{A},$ are the
operators associated to dichotomic observables with values $\left\{
0,1\right\} $. We may try to construct a HVM (or ontic model), as in the
classical case, via a set of functions $\left\{ a\left( \lambda ,A\right)
,b\left( \lambda ,B\right) ,...\right\} $ such that an equation similar to
eq.$\left( \ref{Bellhv}\right) $ holds true. At a difference with the
classical case, quantum mechanics does not predict expectations for all
products of observables (or more correctly, the expectations depend on the
ordering of the operators), but only for products of ``compatible''
observables, that is those whose associated operators commute with each
other. For the sake of clarity I will start with a simple example.

Let us consider a state $\hat{\rho}$ of a quantum system and a set of 3
observables $\left\{ A,B,C\right\} $ each with values $\left\{ 0,1\right\} ,$
represented in quantum mechanics by the projection operators $\left\{ \hat{A}%
,\hat{B},\hat{C}\right\} .$ We assume that $\hat{A}$ commutes with both $%
\hat{B}$ and $\hat{C},$ but $\hat{B}$ does not commute with $\hat{C}.$ (A
physical instance would consists of the observables position, momentum and
spin of a particle, although here we consider for simplicity that the
possible values of every observable are $0$ and $1$). In this case \textit{%
quantum mechanics predicts} the following expectations, to be obtained via
equations similar to eq.$\left( \ref{Bellro}\right) ,$%
\begin{equation}
\left\langle A\right\rangle ,\left\langle B\right\rangle ,\left\langle
C\right\rangle ,\left\langle AB\right\rangle ,\left\langle AC\right\rangle ,
\label{ABC}
\end{equation}
but it does not predict the expectations (involving incompatible
observables, which cannot be measured simultaneously) 
\[
\left\langle BC\right\rangle ,\left\langle ABC\right\rangle . 
\]
We shall assume that commuting (noncommuting) operators correspond to
compatible (incompatible) observables. Therefore the expectations $%
\left\langle A\right\rangle ,\left\langle B\right\rangle ,\left\langle
AB\right\rangle $ may be measured simultaneously in a context, say 1,and the
observables $\left\langle A\right\rangle ,\left\langle C\right\rangle
,\left\langle AC\right\rangle $ may be measured in another context, say 2,
but it is not possible to measure the five expectations eq.$\left( \ref{ABC}%
\right) $ in the same context (otherwise the observables $A,B,C$ would be
compatible). It is trivial to get functions $\left\{ a_{1}\left( \lambda
,A\right) ,b_{1}\left( \lambda ,B\right) \right\} $ giving the expectations $%
\left\langle A\right\rangle ,\left\langle B\right\rangle ,\left\langle
AB\right\rangle $ via eqs.$\left( \ref{Bellhv}\right) .$ This would provide
a HVM (or ontic model) for the context 1. Similarly we may obtain another
HVM for the context 2 via the functions $\left\{ a_{2}\left( \lambda
,A\right) ,c_{2}\left( \lambda ,C\right) \right\} .$ The two HVM toghether
may be called \textit{a single HVM}, that is \textit{noncontextual
(contextual)} if $a_{1}\left( \lambda ,A\right) =$ $a_{2}\left( \lambda
,A\right) $ (if $a_{1}\left( \lambda ,A\right) \neq $ $a_{2}\left( \lambda
,A\right) ).$ The example shows that noncontextual are a particular kind of
HVM. Studying whether they are possible for all pairs $\left\{ \hat{\rho}%
,P\right\} $ of a state, $\hat{\rho}$, and a set of observables, $P$, of a
physical system is the aim of the next subsection.

It may be realized that a noncontextual HVM corresponds to a \textit{%
complete random variables representation} of the pair $\left\{ \hat{\rho}%
,P\right\} ,$ as defined in the mathematical section 2. Indeed the knowledge
of all expectations like eqs.$\left( \ref{Bellhv}\right) $ determines a
joint probability distribution of all observables. I stress that the
commutativity of all operators with each other is a \textit{sufficient}
condition for the existence of a noncontextual HVM, but it is not \textit{%
necessary}.

\subsection{Contextual hidden variables. Kochen-Specker theorem}

\emph{In a Hilbert space of dimension 3 or more, noncontextual hidden
variables theories are not possible in general. }

The result is called \emph{Kochen-Specker theorem}\cite{Kochen-Specker} for
the authors who proved it in 1967, although it had been proved independently
in 1966 by Bell\cite{Bell66}, \cite{Mermin}. (It may be shown that for
spaces of dimension 2 noncontextual HVM are always possible). The original 
\emph{Kochen-Specker }proof will be given here, but once we know that
noncontextual HVM correspond to complete random variables representations, a
simple proof consists of finding an example violating a Bell inequality.
Such examples are given by eqs.$\left( \ref{CH}\right) $ or $\left( \ref
{CHSH}\right) .$ In fact in that case it is not possible a complete random
variables representation, and therefore a noncontextual HVM. The proof of
the Kochen-Specker theorem via Bell inequalities, here presented, has the
advantage of suggesting empirical tests. Indeed experiments have been
performed showing the violation of noncontextual HVM. In particular
empirical Bell tests refute noncontextual HVM (and even the particular class
of noncontextual that are also local HVM if the experiments block the
locality loophole).

What makes necessary the use of contextual HVM is the fact that the
operators associated to observables in quantum theory do not always commute
with each other. But I shall point out that sometimes a noncontextual HVM is
possible for systems where not all such operators commute. Indeed \textit{%
the possibility of noncontextual HVM is not a property of the system but it
depends also on the quantum state.} Thus ``noncontextual'' does not imply
that the joint probability distribution of all observables may be measured
in the same context. Indeed, for some states, noncontextual HVM are possible
even if not all operators commute.

Noncontextual ontic models are always possible in classical physics, whence
the impossibility of such models is a signal of quantum behaviour. Thus the
Bell inequalities may be used, and have been used, in order to discriminate
quantum vs. classical phenomena. If a Bell inequality is violated the
phenomenon cannot be interpreted via noncontextual HVM and it should be
considered specifically quantum.

\subsection{Bell's theorem}

Both in\textit{\ }classical mechanics and in ordinary life the correlations
between distant systems are assumed to derive from a common past. For
instance the similarity between twins (possibly living in different cities)
is a correlation between distant bodies. It is an obvious consequence of the
common origin, which might be formalized stating 
\begin{equation}
\left\langle AB\right\rangle =\int f\left( \lambda \right) a\left( \lambda
,A\right) b\left( \lambda ,B\right) d\lambda ,  \label{twins}
\end{equation}
where $A$ and $B$ label some feature of the twins, e. g. colour of the eyes,
and $f\left( \lambda \right) $ represents the (common) genetic code. But
correlations cannot be interpreted in terms of eqs.$\left( \ref{twins}%
\right) $ or $\left( \ref{Bellhv}\right) $ whenever a Bell inequality is
violated. Hence a fundamental consequence of Bell's work is to show that 
\textit{in nature there might be} \textit{correlations between distant
bodies not deriving from a common past; this would be the case if the
correlations violated a Bell inequality. }The result is a consequence of the
laws of (standard) probabilities and it is therefore independent of the
existence of quantum mechanics.

Quantum mechanics enters because it predicts the existence of such
correlations, a result known as ``Bell's theorem''. Bell introduced hidden
variables models, called local, that are partially contextual. He stressed
that contextual HVM are not too strange provided contextuality means
influence of the context (e. g. the measuring equipment) \emph{near the
system} under study. However it would be rather strange if the influence is
at a distance, as in the experiments of the EPR type\cite{EPR}. In those
experiments two particles are produced in a source, each particle travels in
a different direction and when they are far appart two different agents,
Alice and Bob, measure their respective properties. In this case it is
unplausible to assume that there may be an influence of the context of Alice
(Bob) on the result of the measurement made by Bob (Alice). But those
influences should exist if a Bell inequality is violated by the four
expectations $\left\langle A_{1}B_{1}\right\rangle ,\left\langle
A_{1}B_{2}\right\rangle ,\left\langle A_{2}B_{1}\right\rangle ,\left\langle
A_{2}B_{2}\right\rangle ,$ where $A_{j}$ ($B_{k}$) are observables measured
by Alice (Bob).

It may be realized that the relevance of Bell's theorem is greater than just
to refute a class of hidden variables theories of quantum mechanics. It
proves that \emph{local realistic models} of natural phenomena are not
compatible with quantum mechanics. Indeed Bell himself reinterpreted the
correlation between two measurements, one by Alice the other one by Bob
within relativity theory\cite{Bell87}. To do that he considered the set of
variables $\lambda $ to be the union of two sets, $\lambda _{a}$ and $%
\lambda _{b}$, consisting each of\ all events in the past light cone of the
measurement performed by Alice and Bob, respectively. Therefore the
violation of a Bell inequality with space-like separated measurements, in
the sense of relativity theory, would imply that the correlation does not
derive from the intersection of the said past light cones.

Thus Bell's theorem seems to prove the incompatibility of quantum mechanics
with relativity theory. The contradiction looks dramatic, but most authors
think that there is no real contradiction because quantum mechanics does not
allow sending superluminal signals from Alice to Bob (or from Bob to Alice).
Actually for many authors what is proved by Bell's theorem is the
nonexistence of hidden variables or maybe the unavoidability of an
instrumentalistic interpretation of quantum mechanics, excluding realistic
interpretations. For this reason it is frequent to state Bell's theorem as
proving the incompatibility between \textit{local realism} and quantum
mechanics.

In view of Bell's theorem many experiments have been performed during the
latter forty years in order to test empirically local realism against
quantum mechanics. The experiments confirmed quantum mehcanics but, due to
the existence of loopholes, they were unable to refute local realism. Only
recently has been possible to perform loophole-free experiments\cite{Hensen}%
, \cite{Shalm}, \cite{Giustina}.

\subsection{Entlanglement, the characteristic trait of quantum mechanics}

One of the main difficulties for a realistic understanding of the quantum
formalism derives from the phenomenon of entanglement. Known since the early
days of quantum mechanics it was crucial in the EPR argument mentioned above%
\cite{EPR}. Its relevance was emphasized by Schr\"{o}dinger in a celebrated
paper\cite{Schrodinger}. He stated that entanglement ``is not one but 
\textit{the} characteristic trait of quantum mechanics''. The problem is
that getting a clear physical picture of entanglement is difficult. In fact,
its standard definition is not made in terms of physical concepts, but
requires abstract mathematics:\ ``\textit{Several physical systems are
entangled if their joint wavefunction (or statevector) cannot be written as
the product of wavefunctions}, \textit{one for each system}''. As a
consequence the common wisdom is that entanglement cannot be pictured in
terms of physical models. Actually a physical interpretation has been
proposed that considers entanglement as a correlation between the
fluctuations of the vacuum fields acting on different systems (maybe
separated at a long distance)\cite{FOS}. However this interpretation is not
popular.

One of the difficulties put by the phenomenon of entanglement may be seen
with the following argument. Let us consider an entangled (singlet) state of
two spin-1/2 particles. It may be represented (ignoring normalization) in
the form 
\begin{equation}
\Psi =\left( \uparrow \downarrow -\downarrow \uparrow \right) ,  \label{4.0}
\end{equation}
which may be expressed in words saying that either electron \#1 has spin up
and electron \#2 spin down or electron \#1 has spin down and electron \#2
spin up. The strange fact is that the state eq.$\left( \ref{4.0}\right) $ is
pure, which is interpreted as a complete knowledge of the whole system, and
nevertheless we have no information about the spin of every electron. We
only know that there are 50-50\% probabilities for the two possibilities, up
and dawn, and in fact these would be the probabilities of the results if
measurements were performed. That is, according to the standard quantum view
we have complete information about the whole two-spin system but no
information about every part. (It is like a student claiming to have
complete knowledge of a whole subject matter but no knowledge at all about
every lesson). This is paradoxical because in ordinary life \textit{complete
information just means information about every part.} This statement is true
not only in classical physics but in all sciences where standard
(Kolmogorov) probability theory is used, e.g. in economics or biology. The
fact may be put in quantitative form in terms of ``entropy'', a measure of
our ignorance (our lack of information). In fact in a system consisting of
several parts the classical (Shannon) entropy fulfils the inequalities 
\begin{equation}
S_{j}\leq S\leq \sum_{j}S_{j},  \label{4.7}
\end{equation}
where $S_{j}$ is the entropy of subsystem $j$ and $S$ the entropy of the
whole system. The latter inequality is called subadditivity and it is not
too relevant here. The former means that \textit{our lack of information
about a system cannot be smaller than about one of the parts}. This is
violated in quantum theory, where the standard measure of ignorance is von
Neumann entropy\cite{von Neumann}, defined by 
\begin{equation}
S=-Tr\left( \hat{\rho}\log \hat{\rho}\right) ,S_{j}=-Tr\left( \hat{\rho}%
_{j}\log \hat{\rho}_{j}\right) .  \label{S}
\end{equation}
In fact for the state eq.$\left( \ref{4.0}\right) $ the von Neumann entropy
fulfils 
\[
S=0,S_{1}=S_{2}=\log 2. 
\]
It is possible to show that the violation of the information inequality,
former eq.$\left( \ref{4.7}\right) ,$ is closely related to the violation of
a Bell inequality\cite{Santos4}. Indeed a necessary condition for both is
that the quantum state is entangled.

A naive solution to the information problem would be assuming that the
quantum statevector eq.$\left( \ref{4.0}\right) $ represents a statistical
mixture (of the two possibilities $\uparrow \downarrow $ and $\downarrow
\uparrow $ ). However this assumption poses well known difficulties because
in quantum physics a linear combination like eq.$\left( \ref{4.0}\right) $
has \textit{empirically testable} differences with a mixture. Of course in
hidden variables models (possibly contextual) the von Neumann entropy eq.$%
\left( \ref{S}\right) $ is no longer the appropriate measure of information.

Entanglement is quite common in quantum physics. For instance in
many-electron systems (atoms, molecules and solid bodies) the Pauli
principle gives rise to entanglement. The fact could be illustrated with the
statevector of a two-electron atom. If the total spin is zero (ignoring
nuclear spin) the joint statevector might be written 
\begin{equation}
\Psi \left( 1,2\right) =\psi \left( \mathbf{r}_{1},\mathbf{r}_{2}\right)
\left( \uparrow \downarrow -\downarrow \uparrow \right) ,  \label{4.1}
\end{equation}
where $\psi \left( \mathbf{r}_{1},\mathbf{r}_{2}\right) $ usually does not
have the form of a product of one-electron wavefunctions. In this case eq.$%
\left( \ref{4.1}\right) $ represents an entangled state involving position
and spin. However entanglement is most relevant when the systems are
separated, as stressed in the EPR paper\cite{EPR}.

We see that entanglement is a correlation, but a strange correlation
different from the classical one by the fact that it may violate a Bell
inequality. Indeed the close connection of entanglement with the Bell
inequalities derives from the fact that \textit{entanglement is, in addition
to necessary, a sufficient condition for the existence of a Bell inequality
violated by (ideal) quantum predictions}\cite{Gisin}.

An interesting question related to entanglement appears if quantum mechanics
is universally valid, which is the common wisdom today. (Thus classical
theories are seen as approximations to quantum theories, the passage from
quantum to classical being an active subject of research in recent years).
If quantum mechanics is universally valid and \textit{we accept the
unrestricted validity of the superposition principle} then all statevectors
correspond to physical states of the system. This assumption leads to
bizarre consequences in the macroscopic domain as was emphasized by
Schr\"{o}dinger\cite{Schrodinger} with his celebrated \textit{cat} example.
It consists of a cat plus a radiactive atom such that the cat is killed when
the atom decays. If every vector of the Hilbert space corresponds to a state
of the system, then there exists amongst others a state represented by the
linear combination 
\begin{mathletters}
\begin{equation}
\frac{1}{\sqrt{2}}(\mid livecat,1\rangle +\mid deadcat,0\rangle ),
\label{4.5}
\end{equation}
where the first term represents the cat alive with the atom excited and the
second term the cat dead with the atom in the ground state.

It is obvious that the description of the state of a macroscopic system
cannot be made \textit{in practice} associating to it a statevector and that
a density operator is a more appropriate description\textit{.} The usual
solution to the difficulty is to take into account that a cat, as any
macroscopic system, cannot be isolated from the environment. Therefore the
state of the cat should not be represented by a pure state like eq.$\left( 
\ref{cat}\right) ,$ but by a reduced density operator resulting from taking
the partial trace over the degrees of freedom of the environment. The
reduced density operator might be interpreted as incomplete information,
with some proability $P$ that the cat is alive and the probability $1-P$
that she is dead. This approach to macroscopic quantum systems is named 
\textit{decoherence induced by the environment}, and it leads from a pure
state of a large system to a \textit{reduced density operator} for any
subsystem.The density operator is named an \textit{improper mixture,}
meaning that it is different from the mixture corresponding to an isolated
sysem prepared in a state without complete control of all relevant
parameters (complete control would lead to a pure state). However
decoherence theory just translates the problem from the cat to the universe.
Indeed it is difficult to stablish a rigorous spatial limit to the
environment and if no limit is put we should consider the whole universe.
Thus the decoherence approach is closely connected to the many worlds
interpretation of quantum mechanics \cite{MWI}. In this case we should
assume that, after the evolution of the cat interacting with the
environment, the universe would consist of a linear combination of
statevectors everyone with a copy of the atom, the cat and everything else
including ``me'' (the observer). For many people a linear combination of two
terms, one of them with ``me'' seing a dead cat and another one with ``me''
seing an alive cat is even more bizarre than a linear combination of living
and dead cat. In spite of this the ``wavefuction of the universe'' has been
advocated by respected physicists.

The Schr\"{o}dinger cat example is also relevant for the discussion of
measurement in quantum theory. In fact a measurement amounts to an
interaction of a system (usually small) with a measuring device (usually
macroscopic). If the state of system plus device is represented by a
wavefunction, then after the interaction the composite system will be
entangled. Indeed in the Schr\"{o}dinger example the cat may be taken as a
measuring device for the state of the atom. The measurement problem has been
crucial for the interpretation of quantum mechanics, but a more detailed
discussion is beyond the scope of this paper (see e.g. \cite{FOS}).

\end{mathletters}

\end{document}